\documentclass[american,english]{IEEEtran}
\usepackage[T1]{fontenc}
\usepackage[latin9]{inputenc}
\usepackage{color}
\usepackage{units}
\usepackage{mathtools}
\usepackage{amsmath}
\usepackage{amssymb}
\usepackage{cancel}
\usepackage{mathdots}
\usepackage{stmaryrd}
\usepackage{stackrel}
\usepackage{graphicx}
\PassOptionsToPackage{version=3}{mhchem}
\usepackage{mhchem}
\usepackage{esint}

\makeatletter
\usepackage{colortbl}
\usepackage{cite}

\usepackage[labelsep=endash]{caption}
\usepackage{nomencl}
\usepackage{flushend}
\usepackage{stfloats}
\usepackage[margin=8pt,font=footnotesize,labelfont=md,labelsep=colon]{caption}

\usepackage{multicol}
\usepackage{float}
\usepackage{lipsum}
\usepackage[figurename=Fig.]{caption}

\makeatother

\usepackage{babel}
\begin{document}
\title{Physical Layer Security in Vehicular Communication Networks in the
Presence of Interference}
\author{\textcolor{black}{\normalsize{}Abubakar U. Makarfi}\textit{\textcolor{black}{\normalsize{}$^{1}$,
}}\textcolor{black}{\normalsize{}Rupak Kharel}\textit{\textcolor{black}{\normalsize{}$^{1}$,}}\textcolor{black}{\normalsize{}
Khaled M. Rabie}\textit{\textcolor{black}{\normalsize{}$^{2}$}}\textcolor{black}{\normalsize{},
Omprakash Kaiwartya$^{3}$, Galymzhan Nauryzbayev}\textit{\textcolor{black}{\normalsize{}$^{4}$}}\textcolor{black}{\normalsize{}}\\
\textcolor{black}{\normalsize{}$^{1}$Department of Computing and
Mathematics, Manchester Metropolitan University, UK}\\
\textit{\textcolor{black}{\normalsize{}$^{2}$}}\textcolor{black}{\normalsize{}Department
of Engineering, Manchester Metropolitan University, UK}\\
\textcolor{black}{\normalsize{}$^{3}$School of Science and Technology,
Nottingham Trent University, UK}\\
\textit{\textcolor{black}{\normalsize{}$^{4}$}}\textcolor{black}{\normalsize{}School
of Engineering and Digital Sciences, Nazarbayev University, Astana,
Kazakhstan}\\
\textcolor{black}{\normalsize{}Emails:\{a.makarfi, r.kharel, k.rabie\}@mmu.ac.uk;
omprakash.kaiwartya@ntu.ac.uk; galymzhan.nauryzbayev@nu.edu.kz}}

\maketitle
\selectlanguage{american}%
\textcolor{black}{\thispagestyle{empty}}
\selectlanguage{english}%
\begin{abstract}
This paper studies the physical layer security of a vehicular communication
network in the presence of interference constraints by analysing its
secrecy capacity. The system considers a legitimate receiver node
and an eavesdropper node, within a shared network, both under the
effect of interference from other users. The double-Rayleigh fading
channel is used to capture the effects of the wireless communication
channel for the vehicular network. We present the standard logarithmic
expression for the system capacity in an alternate form, to facilitate
analysis in terms of the joint moment generating functions (MGF) of
the random variables representing the channel fading and interference.
Closed-form expressions for the MGFs are obtained and Monte-Carlo
simulations are provided throughout to validate the results. The results
show that performance of the system in terms of the secrecy capacity
is affected by the number of interferers and their distances. The
results further demonstrate the effect of the uncertainty in eavesdropper
location on the analysis.
\end{abstract}

\begin{IEEEkeywords}
\textcolor{black}{Physical layer security, secrecy capacity, interference,
double Rayleigh fading channels},\textcolor{black}{{} moment generating
functions, vehicular communications.}
\end{IEEEkeywords}

\section{\textcolor{black}{Introduction}}

Recent trends in wireless communications have brought significant
research interest in physical layer security of wireless systems.
This assertion is certainly supported by the robust amount of literature
generated on the subject, covering wide aspects of wireless communications.
For instance, performance of secure cooperative systems over correlated
Rayleigh fading channels was studied in \cite{7506610}, while the
secrecy outage probability over correlated composite Nakagami-$m$/Gamma
fading and the secrecy capacity in the presence of multiple eavesdroppers
over Nakagami-$m$ channels were considered in \cite{sop_nak} and
\cite{sec_cap_nakagami} respectively. Furthemore, the secrecy capacity
in generalised fading has been studied over $\kappa$-$\mu$ shadowed
fading channels \cite{psl}, over $\alpha$-$\mu$/$\kappa$-$\mu$
and $\kappa$-$\mu$/$\alpha$-$\mu$ fading channels \cite{Bhargavsc}
and over Fisher-Snedecor $\mathcal{F}$ composite fading channel \cite{fisher_badaneh}.
Physical layer security has also been invetigated in power-line communication
(PLC) systems for cooperative relaying \cite{khaled_pls_plc} and
over correlated log-normal cooperative PLC channels \cite{pls_lognormal},
as well as for RF energy harvesting in multi-antenna relaying networks
\cite{pls_eh_relay}, to mention a few.

With respect to vehicular communications, physical layer security
of double Rayleigh fading for vehicular communications was studied
in \cite{psl_v2v}, while a relay-assisted mobile network in the double
Rayleigh channel was studied in \cite{relay_dbl_ray}. The double-Rayleigh
channel has particularly been shown, from experimental measurements,
to be a more appropriate model for the high mobility of nodes in a
vehicular network, rather than the more common Rayleigh or Nakagami-$m$
distributions \cite{akki_dbl_ray,dbl_ray2}. The significance of investigating
the physical layer security in a vehicular network is crucial due
to rapid advancements towards autonomous vehicles and smart/cognitive
transportation networks to minimise the risk from compromise. It has
however been observed that the effect of interference on the physical
layer security has received much less attention, even though interference
is inherent within shared networks \cite{interferenceIoT_kaiwartya}
and has been shown to affect the secrecy performance. For example,
in \cite{CR_psl_int} the effect of interference on the secrecy capacity
of a cognitive radio (CR) network was examined, while in \cite{psl_int},
the effect of interference on secrecy outage probability was considered
in a Rayleigh faded channel. Studying interference is of interest
in vehicular networks because the IEEE 1609.4 standard suggests selecting
the least congested channel for data transmission \cite{fuzzyVCPS}.
Additionally, in such highly mobile amd dense networks, the received
signal-to-interference-and-noise ratio (SINR) is routinely used as
the channel quality measure along with geocasting and other geometry-based
localisation techniques \cite{geometryGPSoutage_ompra,measuresVCPS}.

From the aforementioned, this study presents two major contributions.
First, we simplify the capacity analysis of the system under interference
constraints by expressing the logarithm of the SINR in a form that
presents the random variables (RVs) as a linear sum in an exponent.
This allows easier analysis in terms of the joint MGFs of the RVs.
We then employ the transformation, to obtain closed-form expressions
for the moment generating function (MGF) of the joint interference
and fading channel to facilitate the analysis of various system parameters.
We evaluate the joint effects of interference and fading on the secrecy
capacity of the system in the presence of an eavesdropper. In particular,
we consider a double Rayleigh fading channel, which has been shown
from the literature to aptly capture the channel characteristics of
vehicular communication networks \cite{akki_dbl_ray,dbl_ray2}. The
effect of the uncertainty of the eavesdropper location was taken into
account, because this parameter has been shown to greatly affect the
analysis \cite{pls_eve_uncertain}, but has been missing from most
analysis on the subject. However, unlike \cite{pls_eve_uncertain}
where the secrecy outage probability over Rayleigh fading was studied,
in this paper we study the secrecy capacity over a double Rayleigh
channel, then in order to account for the uncertainty in the eavesdropper
location, we model its distance as a random variable (RV). It is worth
noting that to the best of our knowledge, in previous literature on
physical layer security, only the effect of the channel fading or
the interference, but not both have been considered for highly mobile
vehicular communications network. Monte Carlo simulations are provided
to verify the accuracy of our analysis. The results show that the
performance of the system in terms of the secrecy capacity is impacted
by the presence of interfering nodes. The results further demonstrate
the effect of uncertainty in eavesdropper's location on the analysis.

The paper is organised as follows. In Section \ref{sec:Sys-Model},
we describe the system under study. Thereafter, in Section \ref{sec:Perf-anal},
we derive expressions for efficient computation of the secrecy capacity
of the network and derive the MGFs of the signal-to-interference and
noise ratio (SINR) in closed-form. Finally, in Sections \ref{sec:Results}
and \ref{sec:Conclusions}, we present the results and outline the
main conclusions.

\section{System Model\label{sec:Sys-Model}}

\begin{figure}[th]
\begin{centering}
\includegraphics[scale=0.4]{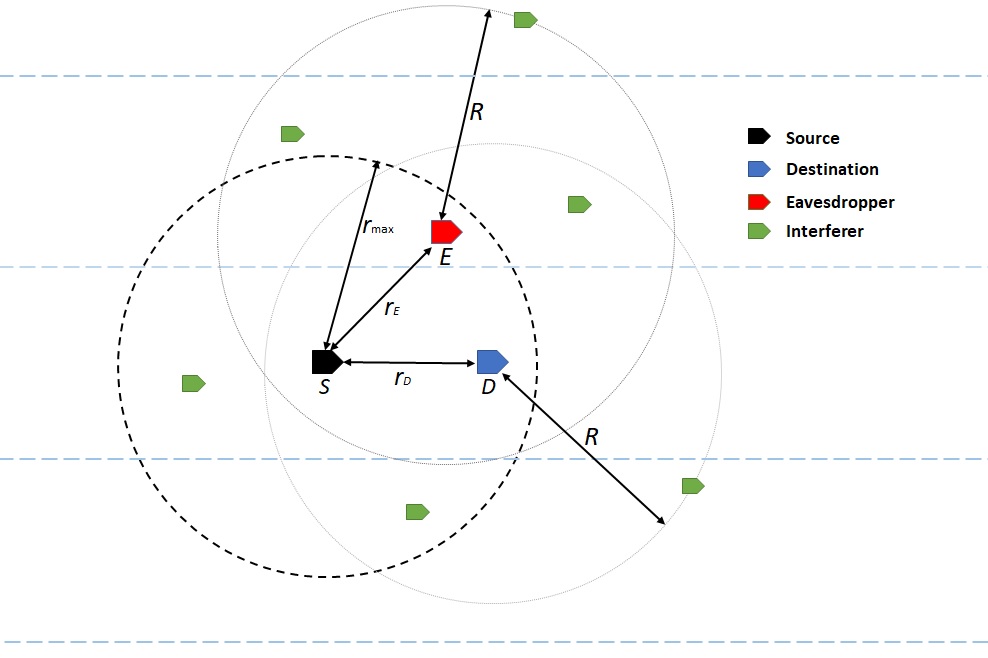}
\par\end{centering}
\caption{General V2V Scenario.\label{fig:sys-mod}}
\end{figure}

Consider a system of nodes operating in a vehicular network. We designate
three nodes of interest: the information source vehicle ($S$), the
information destination vehicle ($D$) and a passive eavesdropper\footnote{Passive eavesdropper in the sense that the node only intercepts the
information, but makes no attempt to actively disrupt, such as through
jamming.} vehicle ($E$). The vehicle $S$ transmits information to the desired
vehicle $D$, while $E$ attempts to receive and decode the confidential
information. Furthermore, the presence of other vehicular nodes operating
within the same space and frequency band, results in co-channel interference
to the received signals of $D$ and $E$. Moreover, while $D$ and
$E$ are known to lie within a certain maximum radius $r_{\textrm{max}}$
from $S,$ the precise relative distances of the vehicle-to-vehicle
(V2V) links are unknown during transmission, which is a realistic
assumption for a network of this nature \cite{pls_eve_uncertain,psl_v2v}.

The received signals at $D$ and $E$ are respectively represented
as
\begin{align}
y_{D} & =h_{D}x+{\displaystyle {\displaystyle \sum_{k=1}^{K}h_{D_{k}}}x_{k}+}w_{D},\label{eq:yD}\\
y_{E} & =h_{E}x+{\displaystyle \sum_{k=1}^{K}}h_{E_{k}}x_{k}+w_{E},\label{eq:yE}
\end{align}
where $x$ and $x_{k}$ are the respective transmitted signals by
$S$ and the $k$-th interferer, with powers $P_{s}$ and $P_{k}$.
The terms $w_{D}$ and $w_{E}$ are the respective additive white
Gaussian noise (AWGN) at $D$ and $E$. Without loss of generality,
we denote the power spectral density of the AWGN as $N_{0}$ and equal
at both links. The terms $h_{i}=\sqrt{g_{i}r_{i}^{-\beta}}\quad i\in\left\{ D,E\right\} $
is the channel coefficient from $S$ to the receiving vehicles $D$
and $E,$ where $r_{i}$ is the V2V link distance, $\beta$ is the
path-loss exponent and $g_{i}$ is the channel gain following double
Rayleigh fading \cite{psl_v2v}. As far as the intereferers are concerned,
$K$ denotes the number of interference nodes, while $h_{i_{k}}=\sqrt{g_{i_{k}}r_{i_{k}}^{-\beta}}\quad i\in\left\{ D,E\right\} $
is the channel coefficient between the $k$th interferer at a distance
$r_{k_{i}}$ from the receiving node, and $g_{i_{k}}$ is the $k$-th
interferer chananel gain.

Based on (\ref{eq:yD}) and (\ref{eq:yE}), the instanstaneous SINRs
at $D$ and $E$ are given by
\begin{equation}
\gamma_{D}=\frac{P_{s}\mid h_{D}\mid^{2}}{\sum_{k=1}^{K}P_{k}\mid h_{D_{k}}\mid^{2}+N_{0}},\label{eq:sinr-d}
\end{equation}
and

\begin{equation}
\gamma_{E}=\frac{P_{s}\mid h_{E}\mid^{2}}{\sum_{k=1}^{K}P_{k}\mid h_{E_{k}}\mid^{2}+N_{0}}.\label{eq:sinr-e}
\end{equation}

\section{Secrecy Capacity Analysis \label{sec:Perf-anal}}

In this section, we derive analytical expressions for the secrecy
capacity of the system. The maximum achievable secrecy capacity is
defined by \cite{bloch_cap}
\begin{equation}
C_{s}=\max\left\{ C_{D}-C_{E},0\right\} ,\label{eq:cs-defined}
\end{equation}
where $C_{D}=\log_{2}\left(1+\gamma_{D}\right)$ and $C_{E}=\log_{2}\left(1+\gamma_{E}\right)$
are the instantaneous capacities of the main and eavesdropping links
respectively. The secrecy capacity in (\ref{eq:cs-defined}) can therefore
be expressed as \cite{bloch_cap}
\begin{equation}
C_{s}=\begin{cases}
\log_{2}\left(1+\gamma_{D}\right)-\log_{2}\left(1+\gamma_{E}\right), & \gamma_{D}>\gamma_{E},\\
0, & \gamma_{D}<\gamma_{E}.
\end{cases}\label{eq:cs-defined-2}
\end{equation}

\subsection{Average Secrecy Capacity}

The average secrecy capacity $\overline{C_{s}}$ is given by \cite{Osamah18GlobecomSC}
\begin{align}
\overline{C_{s}} & =\mathbb{E}\left[C_{s}\left(\gamma_{D},\gamma_{E}\right)\right]\nonumber \\
 & =\stackrel[0]{\infty}{\int}\stackrel[0]{\infty}{\int}C_{s}\left(\gamma_{D},\gamma_{E}\right)f\left(\gamma_{D},\gamma_{E}\right)\textrm{d}\gamma_{D}\textrm{d}\gamma_{E},\label{eq:av-sec-cap-1}
\end{align}
where $\mathbb{E}\left[\cdot\right]$ is the expectation operator
and $f\left(\gamma_{D},\gamma_{E}\right)$ is the joint PDF of $\gamma_{D}$
and $\gamma_{E}$. It is worth noting at this point that the average
in (\ref{eq:av-sec-cap-1}) is with respect to $\gamma_{D}$ and $\gamma_{E}$.
However, assuming each SINR term has $L$ RVs, then we would in turn
require at least $L$-fold numerical integrations to average out the
RVs \{$g_{D},g_{D_{1}}\ldots g_{D_{K}},g_{E},g_{E_{1}}\ldots g_{E_{K}},r_{D},$$r_{E},r_{D_{1}}\ldots$$r_{D_{k}}$
and $r_{E_{1}}\ldots r_{E_{k}}$\} contained within each SINR term.
Obtaining a closed-form solution would be at least arduous, if not
impossible. However, the computational complexity of the task is greatly
reduced by adopting the MGF approach, as mentioned earlier.

We commence by expressing the logarithmic function in (\ref{eq:cs-defined})
in an alternate form. Recalling the identity \cite[Eq. (6)]{hamdi_cap_mrc}
\begin{equation}
\textrm{ln}\left(1+x\right)=\stackrel[0]{\infty}{\int}\frac{1}{s}\left(1-e^{-xs}\right)e^{-s}\textrm{d}s,\label{eq:log-id-1}
\end{equation}
and by substituting $x=\gamma_{D}$ in (\ref{eq:log-id-1}), we can
express the instantaneous capacity of the main link as
\begin{equation}
C_{D}=\frac{1}{\textrm{ln}\left(2\right)}\stackrel[0]{\infty}{\int}\frac{1}{s}\left(1-e^{-s\frac{P_{s}\mid h_{D}\mid^{2}}{\sum_{k=1}^{K}P_{k}\mid h_{D_{k}}\mid^{2}+N_{0}}}\right)e^{-s}\textrm{d}s,\label{eq: log-id-2}
\end{equation}
which after an interchange of variables $s=z\left(\sum_{k=1}^{K}P_{k}\mid h_{D_{k}}\mid^{2}+N_{0}\right)$
and some algebraic manipulations, we obtain an expression in the desired
form with the RVs appearing only in the exponent. Thus,
\begin{multline}
C_{D}=\frac{1}{\textrm{ln}\left(2\right)}\stackrel[0]{\infty}{\int}\frac{1}{z}e^{-zN_{0}}\times\\
\left(e^{-z\sum_{k=1}^{K}P_{k}\mid h_{D_{k}}\mid^{2}}-e^{-zP_{s}\mid h_{D}\mid^{2}}e^{-z\sum_{k=1}^{K}P_{k}\mid h_{D_{k}}\mid^{2}}\right)\textrm{d}z,\label{eq:log-id-3-1}
\end{multline}
and after taking the expectation, we obtain the average capacity of
the main link as
\begin{align}
\overline{C_{D}} & =\mathbb{E}\left[\textrm{ln}\left(1+\frac{P_{s}\mid h_{D}\mid^{2}}{\sum_{k=1}^{K}P_{k}\mid h_{D_{k}}\mid^{2}+N_{0}}\right)\right]\nonumber \\
 & =\frac{1}{\textrm{ln}\left(2\right)}\stackrel[0]{\infty}{\int}\frac{1}{z}e^{-zN_{0}}\left(\mathcal{M}_{\phi_{D}}\left(z\right)-\mathcal{M}_{\chi_{D},\phi_{D}}\left(z\right)\right)\textrm{d}z,\label{eq:av-CD-defined}
\end{align}
where $\mathcal{M}_{\phi_{D}}\left(z\right)=\mathbb{E}\left[e^{-z\sum_{k=1}^{K}P_{k}g_{D_{k}}r_{D_{k}}^{-\beta}}\right]$
is the MGF of the cumulative interference at $D$ and $\mathcal{M}_{\chi_{D},\phi_{D}}\left(z\right)=\mathbb{E}\left[e^{-z\left(P_{s}g_{D}r_{D}^{-\beta}+\sum_{k=1}^{K}P_{k}g_{D_{k}}r_{D_{k}}^{-\beta}\right)}\right]$
is the joint MGF of the eavesdropping link and cumulative interference
at $D$. Using similar analysis, the average capacity of the eavesdropper
link can be represented as
\begin{equation}
\overline{C_{E}}=\frac{1}{\textrm{ln}\left(2\right)}\stackrel[0]{\infty}{\int}\frac{1}{z}e^{-zN_{0}}\left(\mathcal{M}_{\phi_{E}}\left(z\right)-\mathcal{M}_{\chi_{E},\phi_{E}}\left(z\right)\right)\textrm{d}z,\label{eq:av-CE-defined}
\end{equation}
where $\mathcal{M}_{\phi_{E}}\left(z\right)$ is the MGF of the cumulative
interference at $E$ and $\mathcal{M}_{\chi_{E},\phi_{E}}\left(z\right)$
is the joint MGF of the eavesdropper link and cumulative interference
at $E$.

From (\ref{eq:av-CD-defined}), (\ref{eq:av-CE-defined}) and (\ref{eq:cs-defined-2}),
the alternate form for the average secrecy capacity in (\ref{eq:av-sec-cap-1})
can be represented as{\small{}
\begin{multline}
\overline{C_{s}}=\frac{1}{\textrm{ln}\left(2\right)}\stackrel[0]{\infty}{\int}\frac{1}{z}e^{-zN_{0}}\left(\mathcal{M}_{\phi_{D}}\left(z\right)-\mathcal{M}_{\chi_{D},\phi_{D}}\left(z\right)\right)\textrm{d}z\\
-\frac{1}{\textrm{ln}\left(2\right)}\stackrel[0]{\infty}{\int}\frac{1}{z}e^{-zN_{0}}\left(\mathcal{M}_{\phi_{E}}\left(z\right)-\mathcal{M}_{\chi_{E},\phi_{E}}\left(z\right)\right)\textrm{d}z.\label{eq:av-sec-cap-2}
\end{multline}
}{\small\par}

From (\ref{eq:av-sec-cap-2}) we observe that the integrals are symmetrical
and differ mainly in the relative locations of $S,$ $D$ and $E.$
Therefore, this signifies the importance of taking into account the
node locations in the analysis. In what follows, we compute the MGFs
presented in (\ref{eq:av-sec-cap-2}).

\subsection{Computation of Moment Generating Functions}

\subsubsection{The MGF $\mathcal{M}_{\phi_{D}}\left(z\right)$}

The MGF of the cumulative interference at $D$ is given by $\mathcal{M}_{\phi_{D}}\left(z\right)=\mathbb{E}\left[e^{-z\sum_{k=1}^{K}P_{k}g_{D_{k}}r_{D_{k}}^{-\beta}}\right]$,
defined by
\begin{align}
\mathcal{M}_{\phi_{D}}\left(z\right) & =\mathbb{E}\left[e^{-z\sum_{k=1}^{K}P_{k}g_{D_{k}}r_{D_{k}}^{-\beta}}\right]\nonumber \\
 & =\prod_{k=1}^{K}\mathbb{\mathbb{E}}\left[e^{-zP_{k}g_{D_{k}}r_{D_{k}}^{-\beta}}\right]\nonumber \\
 & =\prod_{k=1}^{K}\stackrel[g]{}{\int}\stackrel[r]{}{\int}e^{-zP_{k}g_{D_{k}}r_{D_{k}}^{-\beta}}f_{r_{D}}(r)f_{g_{D}}(g)\textrm{d}r_{D}\textrm{d}g_{D}\label{eq:mgf-int-1}
\end{align}
where $f_{g_{D}}(g)$ and $f_{r_{D}}(r)$ are the probability density
functions (PDFs) of the channel gain $g_{D_{k}}$and interferer distances
$r_{D_{k}}$ respectively.

Let us start by defining a special case of the MGF in (\ref{eq:mgf-int-1})
with only the RVs, given by 
\begin{align}
\mathcal{M}_{\psi}\left(z\right) & =\mathbb{E}\left[e^{-zg_{D_{k}}r_{D_{k}}^{-\beta}}\right]\nonumber \\
 & =\stackrel[g]{}{\int}\stackrel[r]{}{\int}e^{-zg_{D}r_{D}^{-\beta}}f_{r_{D}}(r)f_{g_{D}}(g)\textrm{d}r_{D}\textrm{d}g_{D},\label{eq:mgf-special}
\end{align}
then from the generalized cascaded Rayleigh distribution, we can obtain
the PDF of the double Rayleigh channel for $n=2$ in \cite[Eq. (8)]{cascaded_ray}
as
\begin{equation}
f\left(g\right)=\textrm{G}_{0,2}^{2,0}\left(\frac{1}{4}g^{2}\Biggl|\negthickspace\begin{array}{c}
-\\
\frac{1}{2},\negthickspace\frac{1}{2}
\end{array}\negthickspace\right),\label{eq:pdf-dbl-ray}
\end{equation}
where \textcolor{black}{$\textrm{G}_{u,v}^{s,t}\left(x\mid\cdots\right)$
is the Meijer\textquoteright s G-function \cite[Eq. (9.302)]{bookV3}.
The }node distances $r_{D},$ are assumed to be uniformly distributed
within a circular region, with radius $R$ around the receiver, with
a PDF given by \cite{unifiedModel4int_shobowale}

\begin{equation}
f\left(r\right)=\begin{cases}
\frac{2r}{R^{2}}, & 0<r_{D}\leq R,\\
0, & \textrm{otherwise.}
\end{cases}\label{eq:r distribution}
\end{equation}
Upon invoking \cite[Eq. (2.33.10)]{book2} along with some manipulations
(for $\beta>2$),\footnote{Solution found using the method of integration by parts. For $\beta>2$
only the free-space model ($\beta=2$) is excluded. Hence, this constraint
is acceptable for our purpose.} the inner integral in (\ref{eq:mgf-int-1}) resolves to{\small{}
\begin{multline}
\stackrel[0]{R}{\int}e^{-zg_{D}r_{D}^{-\beta}}\frac{2r_{D}}{R^{2}}\textrm{d}r_{D}=\\
\frac{R^{2}e^{-zgR^{-\beta}}-\left(zg\right){}^{\frac{2}{\beta}}\Gamma\left(1-\frac{2}{\beta},zgR^{-\beta}\right)}{R^{2}},\label{eq:r-resolved}
\end{multline}
}where $\Gamma\left(a,b\right)=\int_{b}^{\infty}t^{a-1}e^{-t}dt$
is the upper incomplete Gamma function \cite[Eq. (9.14.1)]{book2}.
Thus, (\ref{eq:mgf-special}) becomes{\small{}
\begin{multline}
\mathcal{M}_{\psi}\left(z\right)=\stackrel[0]{\infty}{\int}\textrm{G}_{0,2}^{2,0}\left(\frac{1}{4}g^{2}\Biggl|\negthickspace\begin{array}{c}
-\\
\frac{1}{2},\negthickspace\frac{1}{2}
\end{array}\negthickspace\right)\times\qquad\qquad\\
\qquad\frac{R^{2}e^{-zgR^{-\beta}}-\left(zg\right){}^{\frac{2}{\beta}}\Gamma\left(1-\frac{2}{\beta},zgR^{-\beta}\right)}{R^{2}}\textrm{ d}g_{D},\label{eq:r-res}
\end{multline}
}which after several manipulations can be expressed as in (\ref{eq:mgf-final})
shown on top of the page, where $_{\phantom{}p}F_{q}\left(\alpha;\beta;z\right)$
is the generalized hypergeometric series \cite[Eq. (9.14.1)]{book2}.

\begin{figure*}[t]
{\small{}
\begin{multline}
\mathcal{M}_{\psi}\left(z\right)=\frac{R^{\beta}\left(2R^{\beta}\left(1-R^{-2\beta}z^{2}\right)^{\frac{1}{2}}+2z\arcsin\left(zR^{-\beta}\right)-\pi z\right)}{2\left(R^{2\beta}-z^{2}\right)\left(1-z^{2}R^{-2\beta}\right)^{\frac{1}{2}}}-\frac{\left(2z\right)^{\frac{2}{\beta}}\Gamma\left(1+\frac{1}{\beta}\right)^{2}\Gamma\left(1-\frac{2}{\beta}\right)}{R^{2}}\\
\qquad+\frac{2\pi\beta z}{R^{\beta}}\left[\frac{_{\,3}F_{2}\left(\frac{3}{2},\frac{3}{2},\frac{1}{2}-\frac{1}{\beta};\frac{1}{2},\frac{3}{2}-\frac{1}{\beta};z^{2}R^{-2\beta}\right)}{4\left(\beta-2\right)}-\frac{z_{\,3}F_{2}\left(2,2,1-\frac{1}{\beta};\frac{3}{2},2-\frac{1}{\beta};z^{2}R^{-2\beta}\right)}{\pi\left(\beta-1\right)R^{\beta}}\right].\label{eq:mgf-final}
\end{multline}
}{\small\par}
\selectlanguage{american}%
\centering{}\rule[0.5ex]{2.03\columnwidth}{0.8pt}\selectlanguage{english}%
\end{figure*}

Therefore, using (\ref{eq:r-res}) in (\ref{eq:mgf-int-1}), we obtain
the desired MGF of interferer statistics at $D$ as
\begin{align}
\mathcal{M}_{\phi_{D}}\left(z\right) & =\prod_{k=1}^{K}\mathcal{M}_{\psi}\left(zP_{k}\right)\nonumber \\
 & =\left\{ \mathcal{M}_{\psi}\left(zP_{K}\right)\right\} ^{K},\label{eq:mgf-D-only}
\end{align}
where the final step in (\ref{eq:mgf-D-only}) was obtained by assuming
identical transmit powers for interferer nodes, such that $P_{1}=P_{2}=\dots=P_{k}=P_{K}.$

\subsubsection{\label{subsec:Joint-MGF-1}The Joint MGF $\mathcal{M}_{\chi_{D},\phi_{D}}\left(z\right)$}

The joint MGF $\mathcal{M}_{\chi_{D},\phi_{D}}\left(z\right)$ is
given by 
\begin{align}
\mathcal{M}_{\chi_{D},\phi_{D}}\left(z\right) & =\mathbb{E}\left[e^{-zP_{s}g_{D}r_{D}^{-\beta}+\sum_{k=1}^{K}P_{k}g_{D_{k}}r_{D_{k}}^{-\beta}}\right]\nonumber \\
 & =\mathbb{E}\left[e^{-zP_{s}g_{D}r_{D}^{-\beta}}e^{-z\sum_{k=1}^{K}P_{k}g_{D_{k}}r_{D_{k}}^{-\beta}}\right]\nonumber \\
 & =\mathcal{M}_{\chi_{D}}\left(z\right)\mathcal{M}_{\phi_{D}}\left(z\right),\label{eq:joint-mgf-2}
\end{align}
where $\mathcal{M}_{\phi_{D}}\left(z\right)$ is given by the expression
(\ref{eq:mgf-D-only}) and $\mathcal{M}_{\chi_{D}}\left(z\right)$
is the MGF of statistics at $D.$ It is worth noting that the system
considered assumes the location of $D$ is known by $S.$ Consequently,
$r_{D}$ is not random and the MGF is conditioned only on the statistics
of the channel. Thus, the expected value for the first MGF in (\ref{eq:joint-mgf-2})
is given by{\small{}
\begin{align}
\mathcal{M}_{\chi_{D}}\left(z\right) & =\mathbb{E}\left[\exp\left(-zP_{s}g_{D}r_{D}^{-\beta}\right)\right]\nonumber \\
 & =\stackrel[0]{\infty}{\int}e^{-zP_{s}g_{D}r_{D}^{-\beta}}\textrm{G}_{0,2}^{2,0}\left(\frac{1}{4}g^{2}\Biggl|\negthickspace\begin{array}{c}
-\\
\frac{1}{2},\negthickspace\frac{1}{2}
\end{array}\negthickspace\right)\textrm{ d}g_{D}.\label{eq:joint-mgf-3}
\end{align}
}{\small\par}

To proceed, we express (\ref{eq:joint-mgf-3}) in a more tractable
form, by re-writing the Meijer G-function in an alternate form. Thus,
upon invoking \cite[Eq. (9.304.3)]{book2}, we get{\small{}
\begin{equation}
\textrm{G}_{0,2}^{2,0}\left(\frac{1}{4}g^{2}\Biggl|\negthickspace\begin{array}{c}
-\\
\frac{1}{2},\negthickspace\frac{1}{2}
\end{array}\negthickspace\right)=gK_{0}\left(g\right),\label{eq:meijer-bessel}
\end{equation}
}where $K_{0}\left(v\right)$ is the modified Bessel function of the
second kind and $0$th order \cite[Eq. (8.407)]{book2}. Using (\ref{eq:meijer-bessel})
and \cite[Eq. (6.621.3)]{book2} along with some basic algebraic manipulations,
we can straightforwardly obtain the desired result as {\small{}
\begin{equation}
\mathcal{M}_{\chi_{D}}\left(z\right)=\frac{4}{3(1+zP_{s}r_{D}^{-\beta})^{2}}{}_{\phantom{}2}F_{1}\left(2,\frac{1}{2},\frac{5}{2},\frac{zP_{s}r_{D}^{-\beta}-1}{zP_{s}r_{D}^{-\beta}+1}\right),\label{eq:mgf-final-2}
\end{equation}
}where $_{\phantom{}2}F_{1}\left(\alpha;\beta;\gamma;z\right)$ is
the Gauss hypergeometric function \cite[Eq. (9.111)]{book2}. Hence,
we obtain $\mathcal{M}_{\chi_{D},\phi_{D}}\left(z\right)$ by substituting
(\ref{eq:mgf-final}), (\ref{eq:joint-mgf-2}) and (\ref{eq:mgf-final-2})
in (\ref{eq:joint-mgf-2}).

\subsubsection{The MGF\textmd{ $\mathcal{M}_{\phi_{E}}\left(z\right)$}}

The MGF of the cumulative interference at $E$ is given by $\mathcal{M}_{\phi_{E}}\left(z\right)=\mathbb{E}\left[e^{-z\sum_{k=1}^{K}P_{k}g_{E_{k}}r_{E_{k}}^{-\beta}}\right]$.
From the definition of the MGF, it can be observed that the computation
of $\mathcal{M}_{\phi_{E}}\left(z\right)$ follows similar analysis
to the interference at $D$. Due to brevity, the analysis will not
be repeated here. Thus, using (\ref{eq:mgf-final}), the desired MGF
is given by
\begin{equation}
\mathcal{M}_{\phi_{E}}\left(z\right)=\left\{ \mathcal{M}_{\psi}\left(zP_{K}\right)\right\} ^{K}.\label{eq:mgf-E-int-only}
\end{equation}

\subsubsection{The Joint MGF $\mathcal{M}_{\chi_{E},\phi_{E}}\left(z\right)$}

The joint MGF $\mathcal{M}_{\chi_{E},\phi_{E}}\left(z\right)$ can
be obtained through similar analysis presented in Sec. \ref{subsec:Joint-MGF-1},
Therefore, from (\ref{eq:joint-mgf-2}), $\mathcal{M}_{\chi_{E},\phi_{E}}\left(z\right)=\mathcal{M}_{\chi_{E}}\left(z\right)\mathcal{M}_{\phi_{E}}\left(z\right)$,
where $\mathcal{M}_{\phi_{E}}\left(z\right)$ is given by (\ref{eq:mgf-E-int-only}).
For the purpose of the system under consideration, the exact location
of $E$ is unknown, but lies at a maximum distance $r_{\max}$ from
$S$. Using the PDFs in (\ref{eq:pdf-dbl-ray}) and (\ref{eq:r distribution}),
we obtain{\small{}
\begin{multline}
\mathcal{M}_{\chi_{E}}\left(z\right)=\mathbb{E}\left[\exp\left(-zP_{s}g_{E}r_{E}^{-\beta}\right)\right]\\
=\stackrel[0]{\infty}{\int}\stackrel[0]{r_{\max}}{\int}e^{-zP_{s}g_{E}r_{E}^{-\beta}}\frac{2r_{E}}{r_{\textrm{max}}^{2}}\textrm{G}_{0,2}^{2,0}\left(\frac{1}{4}g^{2}\Biggl|\negthickspace\begin{array}{c}
-\\
\frac{1}{2},\negthickspace\frac{1}{2}
\end{array}\right)\textrm{d}r_{E}\textrm{d}g_{E}.\label{eq:mgf-eve-def}
\end{multline}
}Comparing (\ref{eq:mgf-eve-def}) and (\ref{eq:mgf-int-1}), shows
that both expressions are similar with maximum distance $R=r_{\textrm{max}}$
and source power $P_{s}$. Thus, using (\ref{eq:mgf-final}), $\mathcal{M}_{\chi_{E}}\left(z\right)=\mathcal{M}_{\psi}\left(zP_{s}|r_{\max}\right)$
and we obtain from (\ref{eq:mgf-final}) and (\ref{eq:mgf-E-int-only})
the expression{\small{}
\begin{alignat}{1}
\mathcal{M}_{\chi_{E},\phi_{E}}\left(z\right) & =\mathcal{M}_{\chi_{E}}\left(z\right)\mathcal{M}_{\phi_{E}}\left(z\right)\nonumber \\
 & =\mathcal{M}_{\psi}\left(zP_{s}|r_{\max}\right)\left\{ \mathcal{M}_{\psi}\left(zP_{K}\right)\right\} ^{K}.\label{eq:joint-mgf-4}
\end{alignat}
}{\small\par}

To recap, the average secrecy capacity is obtained from (\ref{eq:av-sec-cap-2})
by substituting the relevant MGFs from (\ref{eq:mgf-final}), (\ref{eq:joint-mgf-2}),
(\ref{eq:mgf-final-2}), (\ref{eq:mgf-E-int-only}) and (\ref{eq:joint-mgf-4}).

\section{Numerical Results and Discussions\label{sec:Results}}

\begin{figure}[th]
\begin{centering}
\includegraphics[scale=0.65]{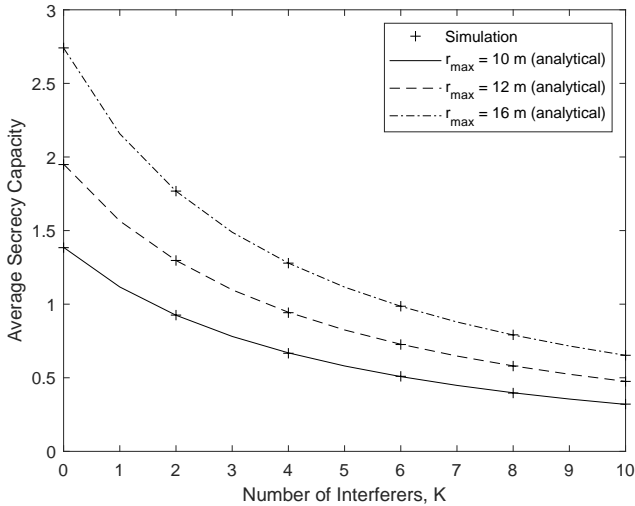}
\par\end{centering}
\caption{Average Secrecy Capacity versus Number of Interferers $K$, with varied
maximum eavesdropper radius $r_{\max}$.\label{fig:cs-vs-k}}
\end{figure}

In this section, we present and discuss some results from the mathematical
expressions derived in the paper. We then investigate the effect of
key parameters on the secrecy capacity of the system. The results
are then verified using Monte Carlo simulations with at least $10^{6}$
iterations. Unless otherwise stated, we have assumed source power
$P_{s}=10$W, interferer transmit power $P_{K}=10$W, $S$-to-$D$
distance $r_{D}=4$m, maximum eavesdropper distance $r_{\max}=10$m,
maximum interferer distance $R=20$m and pathloss exponent $\beta=2.7$.

In Fig. \ref{fig:cs-vs-k}, we present a plot of the average secrecy
capacity against the number of interfering sources in the network.
It can be observed from the results that the number of active interfering
nodes have a negative impact on the secrecy capacity, with the highest
secrecy capacity available when there is no interfering node ($K=0)$
and rapidly reduces with interference. Moreover, this impact can be
effective at various eavesdropper distances from the node, as seen
when $r_{\max}$ is increased. It should be noted that the parameter
$r_{\max}$ is a proxy for the uncertainty of $E$'s location. In
a practical scenario, a vehicle is more likely to know the location
of the node $D$ in which it establishes communication as against
a passive eavesdropper whose presence may not be known. Therefore,
we assume both $D$ and $E$ are always within the radius $r_{\max}$,
while the interferer nodes are restricted by a larger outer radius
of $R.$ Therefore the increased secrecy observed when $r_{\max}$
increases indicates that when $E$ is more likely to be closer to
$S$-to-$D$ V2V link, then the secrecy is compromised, and vice versa.

\begin{figure}[th]
\begin{centering}
\includegraphics[scale=0.65]{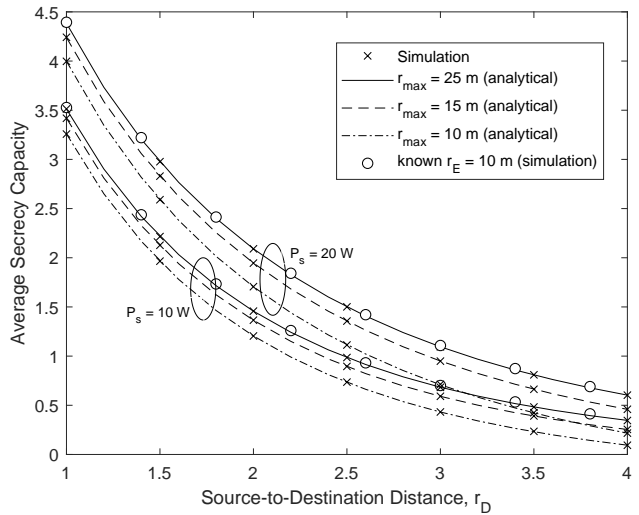}
\par\end{centering}
\caption{Average Secrecy Capacity versus source to destination ($S$-to-$D$
link) for varied source transmit power $P_{s}$ and known eavesdropper
distance $r_{E}$.\label{fig:cs-vs-rd}}
\end{figure}

Fig. \ref{fig:cs-vs-rd}, shows a plot of the average secrecy capacity
against the $S$-to-$D$ distance $r_{D}$, with different values
of $P_{s}$ and $r_{\max}.$ We assume $R=40$m and $5$ interfering
nodes. First, we observe that the average secrecy capacity decreases
as $D$ moves away from $S$, which is expected because the SINR at
$D$ is also decreasing. Next, we see the effect of increasing the
maximum range of $E.$ At the different $P_{s}$ values, we observe
that increasing $r_{\max}$ improves the secrecy capacity because
this means the likely radius of finding $E$ is extended. However,
as $r_{\max}$ is reduced, the secrecy capacity rapidly decreases
and reaches zero approximately when $r_{D}=\nicefrac{1}{2}r_{\max}.$
This shows the significance of the relative locations of $S,$ $D$
and $E.$ To further demonstrate this, we assume a known location
for $E$ and use this distance $r_{E}$ to illustrate the significance
of our result with respect to interferer impact on secrecy. We assume
$r_{E}=10$m and plot the exact secrecy capacity for the V2V network,
through simulations. From Fig. \ref{fig:cs-vs-rd}, it can be observed
that the secrecy capacity for known $r_{E}$ is much superior to the
case when $E$'s location is uncertain. In fact, from our analysis,
it can be seen that the secrecy capacity at $r_{E}=10$m is equivalent
to the secrecy capacity when $r_{\max}=25$m within the region studiesd.
This therefore signifies the importance of taking into account the
uncertainty of the location of $E$, especially for the security analysis
of passive eavesdroppers, when the eavesdropper is unlikely to give
away its position by transmissions.

\begin{figure}[th]
\begin{centering}
\includegraphics[scale=0.65]{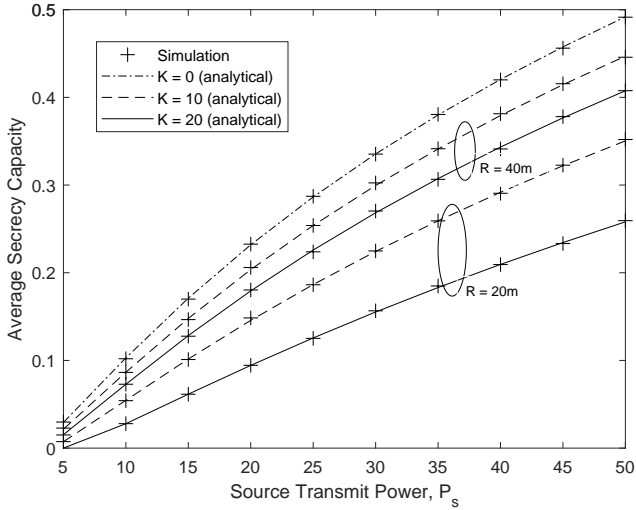}
\par\end{centering}
\caption{Average Secrecy Capacity versus Source Transmit Power $P_{s}$ for
varied interferer numbers $K$ and maximum interferer distances $R.$\label{fig:cs-vs-ps}}
\end{figure}

In Fig. \ref{fig:cs-vs-ps}, we present the average secrecy capacity
with respect to $P_{s}$ for different number of interfering nodes
$K$ and maximum interferer range $R$. As expected, the secrecy capacity
increases monotonically with increased $P_{s}$. Furthermore, within
the region investigated, the average secrecity capacity is highest
without any interfering nodes and degrades with more active interfering
nodes, as already demonstrated in the previous figure. Additionally,
it can be observed that for the same number of interferers, increasing
$R,$ improves the secrecy capacity of the system. Given that both
$S$ and $E$ are affected by the interference in the network, then
an increased radius of interferers, reduces the density of interfering
nodes, which in turn improves the secrecy capacity.

\section{Conclusions\label{sec:Conclusions}}

In this paper, we examined the impact of interference on the secrecy
capacity as a key metric for physical layer security of a wireless
vehicular communication network. Due to the nature of the network
and statistics of the SINR, we expressed the capacity in terms of
the MGF of the RVs representing the joint fading and random distances
of interfering nodes, and eavesdropper node. This reduced the complexity
of the solution. We found closed-form expressions for the various
MGFs and used these expressions to calculate the average secrecy capacity
of the system. The close agreement between the analytical and simulated
results clearly indicated the validity of the derived expressions.
The results demonstrated the effect of some key system parameters
such as the distances of the V2V node distances and eavesdropper,
as well as the number and distances of interference nodes. The results
showed the importance of the analysis with respect to considering
uncertainty of the eavesdropper location, by showing a significant
reduction in secrecy capacity when location of eavesdropper is not
known. The results also showed the effect of interfering nodes on
the security of the system, there by further highlighting the importance
of our analysis.


\bibliographystyle{IEEEtran}
\bibliography{bibGC19}

\begin{thebibliography}{10}
\providecommand{\url}[1]{#1}
\csname url@samestyle\endcsname
\providecommand{\newblock}{\relax}
\providecommand{\bibinfo}[2]{#2}
\providecommand{\BIBentrySTDinterwordspacing}{\spaceskip=0pt\relax}
\providecommand{\BIBentryALTinterwordstretchfactor}{4}
\providecommand{\BIBentryALTinterwordspacing}{\spaceskip=\fontdimen2\font plus
\BIBentryALTinterwordstretchfactor\fontdimen3\font minus
  \fontdimen4\font\relax}
\providecommand{\BIBforeignlanguage}[2]{{%
\expandafter\ifx\csname l@#1\endcsname\relax
\typeout{** WARNING: IEEEtran.bst: No hyphenation pattern has been}%
\typeout{** loaded for the language `#1'. Using the pattern for}%
\typeout{** the default language instead.}%
\else
\language=\csname l@#1\endcsname
\fi
#2}}
\providecommand{\BIBdecl}{\relax}
\BIBdecl

\bibitem{7506610}
Y.~Y. {Zu} and K.~{Xiao}, ``Outage performance of secure cooperative systems
  over correlated rayleigh fading channels,'' in \emph{2016 25th Wireless and
  Optical Commun. Conf. (WOCC)}, May 2016, pp. 1--5.

\bibitem{sop_nak}
G.~C. {Alexandropoulos} and K.~P. {Peppas}, ``Secrecy outage analysis over
  correlated composite {Nakagami-$m$ /Gamma} fading channels,'' \emph{IEEE
  Commun. Lett.}, vol.~22, no.~1, pp. 77--80, Jan. 2018.

\bibitem{sec_cap_nakagami}
M.~Z.~I. {Sarkar}, T.~{Ratnarajah}, and M.~{Sellathurai}, ``Secrecy capacity of
  {Nakagami}-$m$ fading wireless channels in the presence of multiple
  eavesdroppers,'' in \emph{Asilomar Conf. Sig. Sys. Comput.}, Nov. 2009, pp.
  829--833.

\bibitem{psl}
M.~{Srinivasan} and S.~{Kalyani}, ``Secrecy capacity of $\kappa-\mu$ shadowed
  fading channels,'' \emph{IEEE Commun. Lett.}, vol.~22, no.~8, pp. 1728--1731,
  Aug. 2018.

\bibitem{Bhargavsc}
N.~{Bhargav} and S.~L. {Cotton}, ``Secrecy capacity analysis for $\alpha $-$\mu
  $ /$\kappa $-$\mu $ and $\kappa $-$\mu $ / $\alpha $-$\mu $ fading
  scenarios,'' in \emph{2016 IEEE 27th Annual Int. Symp. Personal, Indoor, and
  Mobile Radio Commun. (PIMRC)}, Sep. 2016, pp. 1--6.

\bibitem{fisher_badaneh}
O.~S. {Badarneh}, P.~C. {Sofotasios}, S.~{Muhaidat}, S.~L. {Cotton},
  K.~{Rabie}, and N.~{Al-Dhahir}, ``{On the Secrecy Capacity of Fisher-Snedecor
  F Fading Channels},'' in \emph{2018 14th Int. Conf. Wireless Mobile Comput.,
  Netw. Commun. (WiMob)}, Oct. 2018, pp. 102--107.

\bibitem{khaled_pls_plc}
A.~{Salem}, K.~M. {Rabie}, K.~A. {Hamdi}, E.~{Alsusa}, and A.~M. {Tonello},
  ``Physical layer security of cooperative relaying power-line communication
  systems,'' in \emph{2016 Int. Symp. Power Line Commun, and its Applications
  (ISPLC)}, Mar. 2016, pp. 185--189.

\bibitem{pls_lognormal}
A.~{Salem}, K.~A. {Hamdi}, and E.~{Alsusa}, ``{Physical Layer Security Over
  Correlated Log-Normal Cooperative Power Line Communication Channels},''
  \emph{IEEE Access}, vol.~5, pp. 13\,909--13\,921, 2017.

\bibitem{pls_eh_relay}
A.~{Salem}, K.~A. {Hamdi}, and K.~M. {Rabie}, ``{Physical Layer Security With
  RF Energy Harvesting in AF Multi-Antenna Relaying Networks},'' \emph{IEEE
  Trans. Commun.}, vol.~64, no.~7, pp. 3025--3038, Jul. 2016.

\bibitem{psl_v2v}
Y.~{Ai}, M.~{Cheffena}, A.~{Mathur}, and H.~{Lei}, ``{On Physical Layer
  Security of Double Rayleigh Fading Channels for Vehicular Communications},''
  \emph{IEEE Wireless Commun. Lett.}, vol.~7, no.~6, pp. 1038--1041, Dec. 2018.

\bibitem{relay_dbl_ray}
I.~{Dey}, R.~{Nagraj}, G.~G. {Messier}, and S.~{Magierowski}, ``Performance
  analysis of relay-assisted mobile-to-mobile communication in double or
  cascaded {Rayleigh} fading,'' in \emph{2011 IEEE Pacific Rim Conf. Commun.
  Comput. Sign. Process.}, Aug. 2011, pp. 631--636.

\bibitem{akki_dbl_ray}
A.~S. {Akki} and F.~{Haber}, ``A statistical model of mobile-to-mobile land
  communication channel,'' \emph{IEEE Trans. Veh. Tech.}, vol.~35, no.~1, pp.
  2--7, Feb. 1986.

\bibitem{dbl_ray2}
V.~{Erceg}, S.~J. {Fortune}, J.~{Ling}, A.~J. {Rustako}, and R.~A.
  {Valenzuela}, ``Comparisons of a computer-based propagation prediction tool
  with experimental data collected in urban microcellular environments,''
  \emph{IEEE J. Sel. Areas Commun.}, vol.~15, no.~4, pp. 677--684, May 1997.

\bibitem{interferenceIoT_kaiwartya}
L.~{Farhan}, O.~{Kaiwartya}, L.~{Alzubaidi}, W.~{Gheth}, E.~{Dimla}, and
  R.~{Kharel}, ``{Toward Interference Aware IoT Framework: Energy and
  Geo-Location-Based-Modeling},'' \emph{IEEE Access}, vol.~7, pp.
  56\,617--56\,630, 2019.

\bibitem{CR_psl_int}
Z.~{Shu}, Y.~{Yang}, Y.~{Qian}, and R.~Q. {Hu}, ``Impact of interference on
  secrecy capacity in a cognitive radio network,'' in \emph{Proc. IEEE Global
  Commun. (GLOBECOM)}, Dec. 2011, pp. 1--6.

\bibitem{psl_int}
D.~S. {Karas}, A.~A. {Boulogeorgos}, G.~K. {Karagiannidis}, and
  A.~{Nallanathan}, ``Physical layer security in the presence of
  interference,'' \emph{IEEE Wireless Commun. Lett.}, vol.~6, no.~6, pp.
  802--805, Dec. 2017.

\bibitem{fuzzyVCPS}
R.~{Kasana}, S.~{Kumar}, O.~{Kaiwartya}, R.~{Kharel}, J.~{Lloret}, N.~{Aslam},
  and T.~{Wang}, ``Fuzzy-based channel selection for location oriented services
  in multichannel {VCPS} environments,'' \emph{IEEE Internet Things J.},
  vol.~5, no.~6, pp. 4642--4651, Dec 2018.

\bibitem{geometryGPSoutage_ompra}
O.~{Kaiwartya}, Y.~{Cao}, J.~{Lloret}, S.~{Kumar}, N.~{Aslam}, R.~{Kharel},
  A.~H. {Abdullah}, and R.~R. {Shah}, ``{Geometry-Based Localization for GPS
  Outage in Vehicular Cyber Physical Systems},'' \emph{IEEE Trans. Veh. Tech.},
  vol.~67, no.~5, pp. 3800--3812, May 2018.

\bibitem{measuresVCPS}
S.~{Kumar}, U.~{Dohare}, K.~{Kumar}, D.~{Prasad}, K.~N. {Qureshi}, and
  R.~{Kharel}, ``Cybersecurity measures for geocasting in vehicular cyber
  physical system environments,'' \emph{IEEE Internet Things J.}, pp. 1--1,
  2019.

\bibitem{pls_eve_uncertain}
D.~S. {Karas}, A.~A. {Boulogeorgos}, and G.~K. {Karagiannidis}, ``Physical
  layer security with uncertainty on the location of the eavesdropper,''
  \emph{IEEE Wireless Commun. Lett.}, vol.~5, no.~5, pp. 540--543, Oct. 2016.

\bibitem{bloch_cap}
M.~{Bloch}, J.~{Barros}, M.~R.~D. {Rodrigues}, and S.~W. {McLaughlin},
  ``Wireless information-theoretic security,'' \emph{IEEE Trans. Inf. Theory},
  vol.~54, no.~6, pp. 2515--2534, Jun. 2008.

\bibitem{Osamah18GlobecomSC}
O.~S. Badarneh, P.~C. Sofotasios, S.~Muhaidat, S.~L. Cotton, K.~Rabie, , and
  N.~Al-Dhahir, ``On the secrecy capacity of {F}isher-{S}nedecor $\mathcal {F}$
  fading channels,'' in \emph{Proc. IEEE Global Commun. (GLOBECOM)}, May 2018.

\bibitem{hamdi_cap_mrc}
K.~A. {Hamdi}, ``Capacity of {MRC} on correlated {Rician} fading channels,''
  \emph{IEEE Trans. Commun.}, vol.~56, no.~5, pp. 708--711, May 2008.

\bibitem{cascaded_ray}
J.~{Salo}, H.~M. {El-Sallabi}, and P.~{Vainikainen}, ``The distribution of the
  product of independent {Rayleigh} random variables,'' \emph{IEEE Trans.
  Antennas Propag.}, vol.~54, no.~2, pp. 639--643, Feb. 2006.

\bibitem{bookV3}
A.~P. Prudnikov, Y.~A. Brychkov, and O.~I. Marichev, \emph{Integrals, and
  Series: More Special Functions}, Gordon and Breach Sci. Publ., New York,
  1990, vol. 3.

\bibitem{unifiedModel4int_shobowale}
Y.~Shobowale and K.~Hamdi, ``A unified model for interference analysis in
  unlicensed frequency bands,'' \emph{IEEE Trans. Wireless Commun.}, vol.~8,
  no.~8, pp. 4004--4013, Aug. 2009.

\bibitem{book2}
I.~S. Gradshteyn and I.~M. Ryzhik, \emph{Table of Integrals, Series, and
  Products}, 7th ed. Academic Press, Califonia, 2007.

\end{thebibliography}


\end{document}